\documentstyle[aps,psfig,multicol]{revtex}  

\begin{document}

\title{Improving efficiency of supercontinuum generation in photonic crystal 
       fibers by direct degenerate four-wave-mixing}

\author{N.I. Nikolov$^{1,2}$ and O. Bang$^{1}$} 

\address{$^{1}$Informatics and Mathematical Modelling, 
  Technical University of Denmark, DK-2800 Kongens Lyngby, Denmark \\ 
  phone:+45 45 25 30 79, fax:+45 45 93 12 35, e-mail:nin@imm.dtu.dk}
\address{$^{2}$Ris{\o} National Laboratory, Optics and Fluid Dynamics 
  Department, OFD-128 P.O. Box 49, DK-4000 Roskilde, Denmark}

\author{A. Bjarklev$^{3}$}
\address{$^{3}$Research center COM, Technical University of Denmark, 
  DK-2800 Kongens Lyngby, Denmark}

\maketitle 

\begin{abstract}
We numerically study supercontinuum (SC) generation in photonic crystal
fibers pumped with low-power 30-ps pulses close to the zero dispersion
wavelength 647nm. We show how the efficiency is significantly improved 
by designing the dispersion to allow widely separated spectral lines
generated by degenerate four-wave-mixing (FWM) directly from the pump 
to broaden and merge. 
By proper modification of the dispersion profile the generation
of additional FWM Stokes and anti-Stokes lines results in efficient 
generation of an 800nm wide SC. 
Simulations show that the predicted efficient SC generation is more 
robust and can survive fiber imperfections modelled as random 
fluctuations of the dispersion coefficients along the fiber length.
\end{abstract}

\begin{multicols}{2}  

\section{Introduction}

After the first observation of a 200-THz supercontinuum (SC) spectrum of light in bulk glass\cite{AlfShap1,AlfShap2}, much has been
done on the understanding and control of this process\cite{BookAlf}. A target of
numerous experimental and theoretical investigations has been the
improvement of the characteristics and simplification of the technical
requirements for the generation of a SC\cite{BookAlf}. The first experiments in bulk
materials, based on self-phase modulation (SPM), required extremely high
peak powers ($\rm{>10MW}$).

New techniques based on the use of optical fibers as
a nonlinear medium for SC generation allowed lower peak powers to be
used due to the long interaction length and high
effective nonlinearity\cite{Lin,Baldeck,SKoen,SKoen1}. However, the necessity
to operate near the wavelength for zero group velocity dispersion, restricted the SC
generation to the spectral region around and above $\rm{1.3\mu m}$.
The use of dispersion-flattened or dispersion-decreasing fibers as
nonlinear media for SC generation resulted in a flat SC spanning
1400-1700nm\cite{K49,K49_new} and 1450-1650nm\cite{K19}, respectively. The
spectrum was still far from the visible wavelengths and in some cases very sensitive to
noise in the input \cite{K19}.

Photonic crystal fibers (PCF) and tapered fibers overcome these
limitations. The unusual dispersion properties and enlarged effective nonlinearities 
make them a promising tool for effective SC generation \cite{SKoen}. 
PCFs and tapered fibers have similar dispersion and nonlinearity 
characteristics and they have the advantage that their dispersion 
may be significantly modified by a proper design of the cladding 
structure \cite{zdisp,zdisp1,zdisp2}, or by changing the degree of
tapering \cite{PCF1}, respectively. 
Using kilowatt peakpower femtosecond pulses a SC spanning 400-1500nm has 
been generated in a PCF \cite{PCF1} and in a tapered fiber \cite{Tapp}. 
The broad SC was later explained to be a result of SPM and direct 
degenerate four-wave-mixing (FWM) \cite{SC_FWM}.

However, high power femtosecond lasers are not necessary, - SC generation
may be achieved with low-power picosecond \cite{SKoen,SKoen1} and even 
nanosecond \cite{SKoen2} pulses. 
Thus Coen \emph{et al.} generated a broad SC in a PCF using 
picosecond pulses with sub-kilowatt peakpower and showed 
that the primary mechanism was the combined effect of stimulated Raman 
scattering (SRS) and parametric FWM, allowing the Raman shifted 
components to interact efficiently with the pump \cite{SKoen}.

Using 200 fs high power pulses and a 1cm long tapered
fiber, Gusakov has shown that direct degenerate FWM can
lead to ultrawide spectral broadening and pulse
compression\cite{SC_FWM}. We consider how the direct degenerate FWM can 
significantly improve the efficiency of SC generation with
sub-kilowatt picosecond pulses in PCFs, and go one step further by
optimizing the effect through engineering of the dispersion properties
of the PCF. We show that by a proper design of the
dispersion profile the direct degenerate FWM Stokes and anti-Stokes
lines can be shifted closer to the pump, thereby allowing them to
broaden and merge with the pump to form an ultrabroad SC. 
This significantly improves the efficiency of the SC generation, 
since the power in the Stokes and anti-Stokes lines no longer is lost.
In particular we optimize the SC bandwidth by designing the dispersion
profile to generate additional Stokes and anti-Stokes lines around 
which the SC spectrum can broaden.

External perturbations and different types of imperfections lead to
fluctuations of the fiber parameters along the length of the 
fiber. Fluctuations in fiber birefringence\cite{CDPole,PKAWai}, dispersion\cite{MKarl,NKuwaki}, and
nonlinearity\cite{JGarnier,RKnapp} has been investigated to understand their influence on
different regimes of light propagation. As parametric processes require phase matching, the
effectiveness of the FWM could be strongly influenced by random
fluctuations of the dispersion. Indeed Coen \emph{et.al.} in their PCF
experiments with low-pump picosecond pulses at
647nm\cite{SKoen,SKoen1} and nanosecond pulses at 532nm\cite{SKoen2} explains the
absence of frequencies generated by direct degenerate FWM from the
pump, by the large frequency shift and the violation of the required
phase-matching condition due to fiber irregularities. We analyze the
influence of a random change of the dispersion coefficients along
the fiber on the process of SC generation and find that the generation
and merging of the direct degenerate FWM Stokes and anti-Stokes
waves with the pump could be robust enough to survive fiber imperfections, and thus
a significant improvement of the process of SC generation should
indeed be possible in real PCFs.

\section{Theoretical model and fiber data.}

We study numerically the SC generation process using the well known coupled
nonlinear Schr\"{o}dinger (NLS) equations that describe the evolution of
the x- and y-polarization components of the field for pulses with a
spectral width of up to 1/3 of the pump frequency \cite{SKoen,BWood},

\begin{eqnarray}
\label{scgq}
\frac{\partial A_j}{\partial z} & = & -\mu A_j  + i(j-1)\delta\beta A_j 
   + (-1)^j\frac{\Delta}{2}\frac{\partial A_j}{\partial\tau} \\
&& -\frac{i}{2}\sum_{k=2}^7\frac{\beta_k}{k!}\frac{\partial^kA_j}{\partial
   \tau^k} + i\gamma\left( 1+\frac{i}{\omega_p}\frac{\partial}{\partial\tau}
   \right) \left\{\frac{ }{ }  \right. \nonumber \\
&& \left. A_j f_R \int h_R(\tau-s) \left[ |A_j(s)|^2 + |A_{3-j}(s)|^2 
   \right] ds \right. \nonumber\\
&& \left. + (1-f_R)\left[\left( |A_j|^2 + \frac{2}{3}|A_{3-j}|^2\right) A_j 
   + \frac{1}{3}A_j^*A_{3-j}^2 \right]\right\}. \nonumber
\end{eqnarray}
Here the complex fields $A_j = A_j (t,z)$ with $j = \rm{1,2}$
are given by $A_1 = E_x$ and $A_2 = E_y \exp (i\delta \beta z)$, where $E_x$ and $E_y$ are the
envelopes of the real linearly polarized x- and y-components. The time
$\tau = t - z/\overline{v}$ is in a reference frame moving with the
average group velocity $\overline{v}^{\rm{-1}} = (v_x^{\rm{-1}} + v_x^{\rm{-1}})/\rm{2}$, $z$
is the propagation coordinate along the fiber, $\mu$ is the fiber
loss, $\delta \beta = \beta_x - \beta_y = \omega_0 \delta n /c$ is the
phase mismatch due to birefringence $\delta n = n_x - n_y$, and
$\Delta = ( v_x^{\rm{-1}} - v_y^{\rm{-1}})$ is the group velocity mismatch between the
two polarization axes. The propagation constant
$\beta(\omega )$ is expanded to $\rm{8^{th}}$ order around the pump frequency
$\omega_p$ with coefficients $\beta_k$ keeping $\beta_{2-7}$ same for
x- and y-linearly polarized components, $\gamma$ is the effective nonlinearity, $f_R$ is the fractional contribution of the Raman effect, and finally $^*$ denotes
complex conjugation.

This model accounts for SPM, cross-phase-modulation, FWM, and SRS. 
For the Raman susceptibility $h_R$ we include only the parallel component, as the orthogonal
component is generally negligible in most of the frequency regime that we
consider\cite{RamOrt}. The Raman susceptibility is approximated by the expression\cite{AG}:
\begin{equation}
h_R (t) = \frac{\tau_1^2  + \tau_2^2}{\tau_1 \tau_2^2} \exp(-t/\tau_2)\sin(t/\tau_1),
\label{ramq}
\end{equation}
where $\tau_1=\rm{12.2}fs$ and $\tau_2=\rm{32}fs$. Furthermore, $f_R
=\rm{0.18}$ is estimated from the known numerical value of the peak
Raman gain\cite{AG}.

The phase-mismatch for degenerate FWM of two photons at
the pump frequency is: $\Delta\beta = \beta_s + \beta_{as}-2\beta_p + 2\gamma (1-f_R)I_p$\cite{AG}, where $I_p$ is the peak
power. In the frequency domain we have:
\begin{equation}
\Delta\beta = 2(\Omega ^2 \frac{\beta _2}{2!} + \Omega ^4
\frac{\beta _4}{4!} + \Omega ^6 \frac{\beta _6}{6!} + \gamma(1-f_R)
I_p ),
\label{phaseq}
\end{equation}
where $\Omega = \omega_p - \omega_s = \omega_{as}
  -\omega_p$. The gain $g$ of the direct degenerate FWM\cite{SKoen1,AG} is:
\begin{equation}
  g=[(\gamma(1-f_R) I )^{2} - (\Delta \beta /2)^{2}]^{1/2},
\label{gainq}
\end{equation}
where $I$ is the power of the frequency component around which the
degenerate FWM process takes place.

We consider the same PCF and use the same numerical and experimental
data as in \cite{SKoen}, kindly provided by S. Coen. We pump along the
slow axis with $\rm{30ps}$ sech-shaped pulses of $I_{p}=\rm{400W}$ 
peak power and pump wavelength $\lambda_p=\rm{647nm}$. 
The PCF has core area $A_{core}=\rm{1.94\mu m^2}$, birefringence
$\delta n =\rm{1.9\cdot 10^{-6}}$, and nonlinearity $n_2=\rm{3\cdot 
10^{-20}m^2/W}$. We consider six different dispersion profiles, for 
which $\beta_{2-7}$ are given in Table \ref{dispt}, where case d1 
corresponds to the PCF used in \cite{SKoen}. Note that cases d4+d6 
and d5 have two and three sets of Stokes and anti-Stokes waves, 
respectively.

Here we are interested in studying the separate effect of different
dispersion profiles (i.e. different values of $\beta_{2-7}$). We
therefore keep all other fiber and pulse parameters constant and
neglect the frequency dependence of the effective area $A_{eff}$ and
the loss $\mu$. Thus we assume a uniform loss of $\mu =\rm{0.1dB/m}$
and approximate $A_{eff}$ with the core area, so that $\gamma=2
n_2/(\lambda_pA_{core})= \rm{0.15(Wm)^{-1}}$.

\begin{table}[bht]
\caption{\label{dispt}
  Dispersion coefficients $\beta_2\,\rm{[ps^2/km]}$, $\beta_4
  \,\rm{[10^{-5}ps^4/km]}$ and $\beta_6\,\rm{[10^{-10}ps^6/km]}$ 
  for dispersion profiles d1-d6, with corresponding dispersion at
  the pump wavelength $D(\lambda_p)\,\rm{[ps/nm\cdot km]}$, zero 
  dispersion wavelength $\lambda_z\,\rm{[nm]}$ and 
  Stokes $\lambda_s\, \rm{[nm]}$ and anti-Stokes $\lambda_{as}\,
  \rm{[nm]}$ wavelengths. Fixed coefficients: $\beta_3=5.1\cdot 
  10^{-2} \rm{ps^3/km}$, $\beta_5=\rm{1.2\cdot 10^{-7} ps^5/km}$
  and $\beta_7=\rm{1.2\cdot 10^{-13} ps^7/km}$}
\begin{tabular}{|*{8}{r|}}
  case & $\beta_2$ & $\beta_4$ & $\beta_6$ & $\lambda_z$ & 
         $D(\lambda_p)$ & $\lambda_s$ & $\lambda_{as}$\\ 
  \hline\hline
  d1   &  7.0  &  -4.9 & -1.8  & 677 & -31.6  & 1110 & 457 \\
  \hline
  d2   & 14    & -34.4 & -0.04 & 697 & -62.3  &  852 & 521 \\
  \hline
  d3   &  1.0  &  -2.5 & -3.3  & 652 &  -4.5  &  852 & 521 \\
  \hline
  d4   & -0.28 &  0.05 &  0.29 & 647 &   1.3  & 1062 & 465 \\
       &       &       &       &     &        &  852 & 521 \\
  \hline
  d5   & -1.01 &  2.14 & -2.84 & 643 &   4.54 & 1104 & 458 \\
       &       &       &       &     &        &  894 & 507 \\
       &       &       &       &     &        &  751 & 569 \\
  \hline
  d6   & -1.3  & -2.6  & 58.8  & 641 &   5.9  &  800 & 543 \\
       &       &       &       &     &        &  720 & 587 \\
 \end{tabular}
\end{table}

For our numerical simulation, we use the standard second order
split-step Fourier method, solving the nonlinear part with
a fourth order Runge-Kutta method using a Fourier transform forth and
back and the convolution theorem. Except where otherwise stated,
we use $2^{17}$ points in a time window of $T=\rm{236ps}$, 
giving the wavelength window ($\rm{405\,-\,1613nm}$). 
The propagation step is $\Delta z = \rm{43\mu m}$. 
In our longest simulation out to $L=\rm{3.7m}$ the photon number 
is conserved to within $\rm{5\%}$ of its initial value. 
An initial random phase noise seeding of one photon per mode is
included as in \cite{SKoen}.
All the presented spectra have been smoothed over 32 points.

\section{Numerical analysis.}

\begin{figure}[h]
\centerline{\psfig{figure=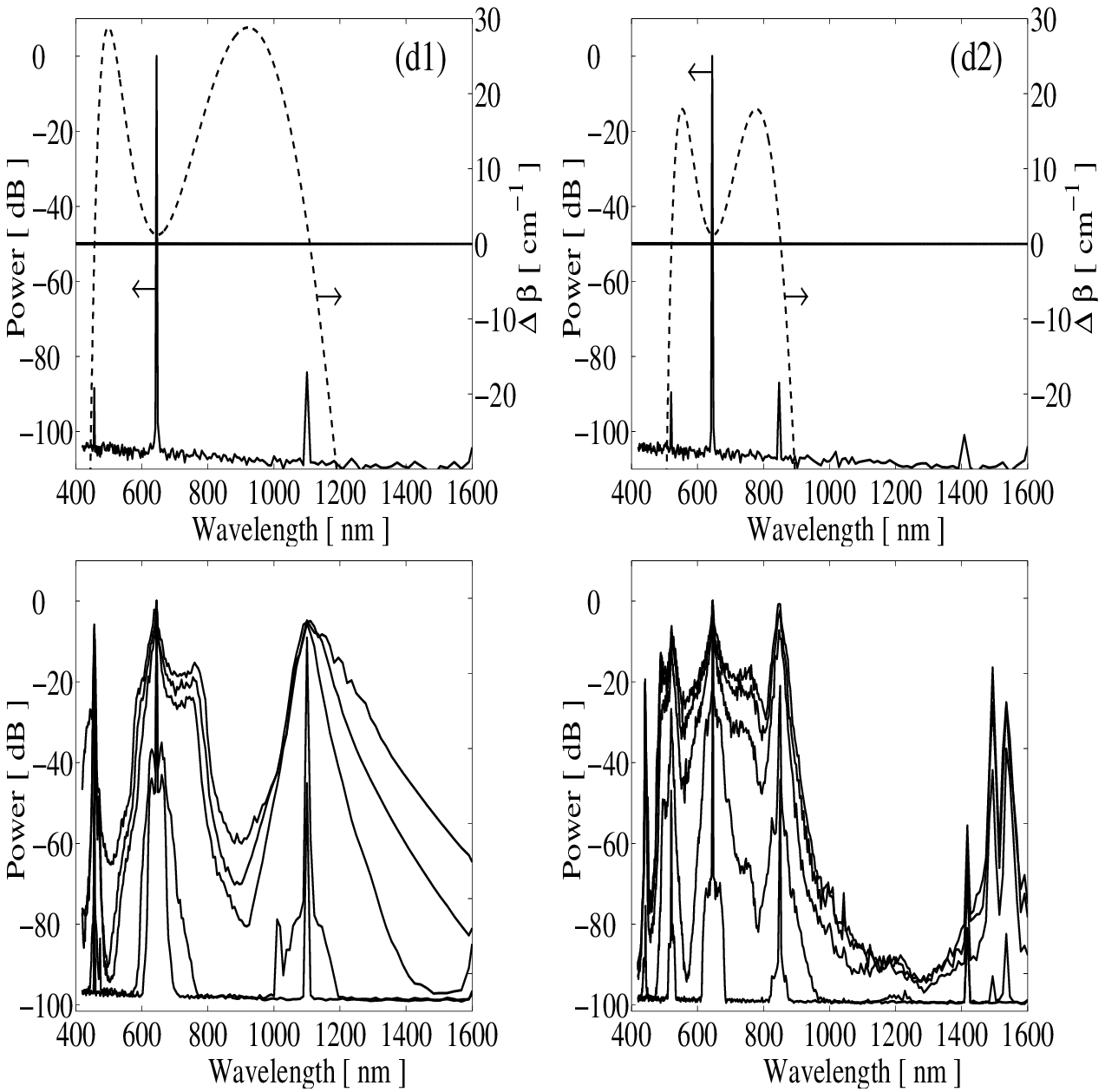,width=1\linewidth}}
\caption{Dispersion profiles d1 (left) and d2 (right). 
  Top row: phase mismatch $\Delta\beta$ for direct degenerate 
  FWM (dashed line) and spectrum at $L$=10cm (solid line). 
  Bottom row: spectrum at $L$=20cm, 30cm, 1m, 2m, and 3.7m 
  (down to up).}
\label{scg1}
\end{figure}

We first simulate SC generation using the same fiber as in \cite{SKoen}, 
i.e., using the dispersion profile d1. Due to our large spectral window 
($\rm{405\,-\,1613nm}$), we see in Fig. \ref{scg1} (left) the emergence 
of direct-degenerate FWM Stokes and anti-Stokes waves at the predicted
wavelengths  $\lambda_s=\rm{1100nm}$ and  $\lambda_{as}=\rm{458nm}$, 
for which the phase matching condition (\ref{phaseq}) is satisfied. 
From the standard expressions given in \cite{AG} we find the maximum direct
degenerate FWM gain ($\rm{g}$) to be twice the maximum SRS gain, which
explains why the FWM Stokes and anti-Stokes components appear before the SRS components.

For a given peak power, the loss and temporal walk-off of the PCF gives 
the maximum distance $L_{max}$
over which nonlinear processes, and thus the SC generation process, are
efficient. From Fig. \ref{scg1} (left) we see that after the FWM Stokes and
anti-Stokes components are generated they broaden in the same way
as the central part of the spectrum around the pump.
The merging of the spectral parts around  $\lambda_{as}$, $\lambda_p$, and
$\lambda_s$ would create an ultra broad spectrum as observed in
tapered fibers with high power femtosecond pulses
\cite{Tapp,SC_FWM}. However, for high power femtosecond pulses the SPM
is the dominant mechanism that leads to broadening and
merging of the Stokes and anti-Stokes lines with the pump. For low
power picosecond pulses the Raman and parametric processes are dominant. 
Thus, in this particular case, the large frequency shift $\sim \rm{193 THz}$ 
of the degenerate FWM sidebands and the narrow degenerate FWM gain bands 
$\sim \rm{2 THz}$ prevent merging of the pump with the Stokes
and anti-Stokes lines to happen within the maximum length $L_{max}$,
i.e., before nonlinear effects become negligible. Indeed it is seen from
Fig.\ref{scg1}, that the spectrum does not change significantly from $\rm{1m}$
to $\rm{3.7m}$.
The power transferred to the Stokes and anti-Stokes lines is in effect lost,
i.e., the SC process is not very efficient.

\begin{figure}[h]
\centerline{\psfig{figure=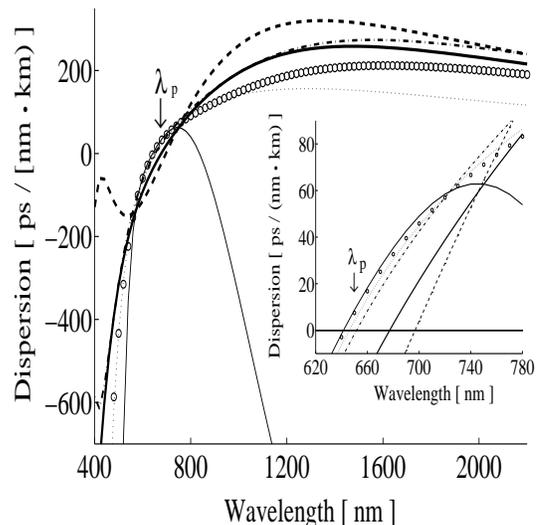,width=0.82\linewidth}}
  \caption{Dispersion profile d1 (thick solid line), d2 (dashed line), 
  d3 (dash-dotted line), d4 (dotted line), d5 (circles), and
  d6 (thin solid line).}
  \label{dispf}
\end{figure}

\begin{figure}[h]
  \centerline{ \vbox{\psfig{figure=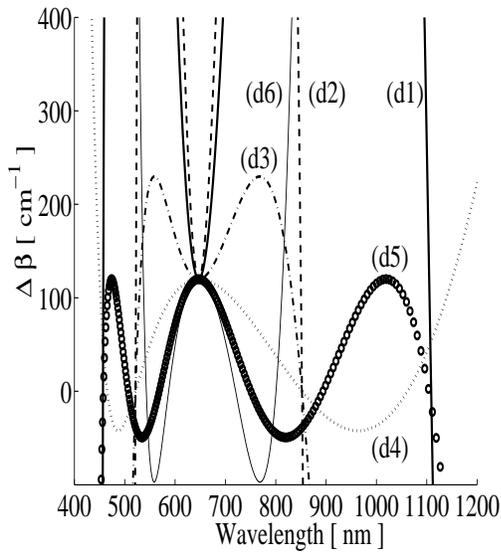,width=0.86\linewidth}}}
  \caption{Phase mismatch $\Delta\beta(\lambda)$ for dispersion profile 
  d1 (thick solid line), d2 (dashed line), d3 (dash-dotted line), 
  d4 (dotted line), d5 (circles), and d6 (thin solid line).}
\label{phasef}
\end{figure}

By increasing the peak power, it would be possible to achieve merging of 
the spectral parts around $\lambda_{as}$, $\lambda_p$, and $\lambda_s$. 
However, our aim is to keep the low peak power fixed and instead achieve
this merging only by engineering the dispersion profile. 
Thus we modify the dispersion profile to adjust Eq.~(\ref{phaseq}) to be 
fulfilled for wavelengths $\lambda_s$ and $\lambda_{as}$ closer to the 
pump $\lambda_p=\rm{647nm}$.  
We do so by modifying $\beta_2$, $\beta_4$, and  $\beta_6$ as listed in
Table \ref{dispt}. The phase-matching condition $\Delta\beta=\rm{0}$
then gives $\lambda_s=\rm{852nm}$ and $\lambda_{as}=\rm{521nm}$ for case 
d2-d3. In case d4-d6 additional Stokes and anti-Stokes waves exist.
The dispersion profiles and phase-mismatch curves corresponding to the
cases in Table \ref{dispt} are shown in Fig. \ref{dispf} and
Fig. \ref{phasef}, respectively. It is important to note that the
curves $\Delta\beta(\lambda)$ have different slope around $\lambda_s$ 
and $\lambda_{as}$ (see Fig. \ref{phasef}).

\begin{figure}[h]
\centerline{\hbox{\psfig{figure=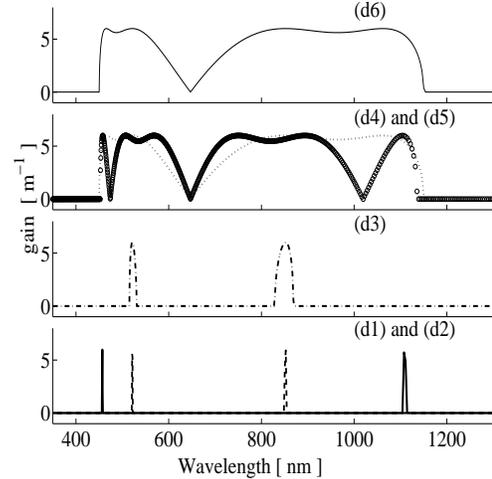,width=0.75\linewidth}}}
  \caption{Degenerate FWM gain $g(\lambda)$ for dispersion profile 
  d1 (thick solid line), d2 (dashed line), d3 (dash-dotted line), 
  d4 (dotted line), d5 (circles), and d6 (thin solid line).}
\label{gainf}
\end{figure}

We first consider only a shift of the Stokes and anti-Stokes lines
closer to the pump and the effect it has on the improvement of the SC
generation. Thus, for the dispersion profile d2, the slope of the
phase mismatch curve around $\lambda_s=\rm{852nm}$ and
$\lambda_{as}=\rm{521nm}$ is kept the same as for case d1. 
It is seen from Fig. \ref{scg1} (right) that such a shift
of the direct degenerate FWM Stokes and anti-Stokes lines closer to
the pump is not enough for a complete merging to take place. This can be
explained by considering the direct degenerate FWM gain $g(\lambda)$ shown
in Fig.~\ref{gainf}. The broadening of the Stokes and anti-Stokes lines
is strongly influenced by the bandwidth of $g(\lambda)$, which is mainly
determined by the slope around $\Delta\beta=0$, i.e., around $\lambda_s$ 
and $\lambda_{as}$. The slope and thus the gain bandwidth is the same in 
case d1 and d2, which explains why the broadening appears to be unchanged.

One way to improve the SC spectrum is to shift the Stokes and anti-Stokes
lines even closer to the pump, keeping the slope of the phase mismatch 
curve around them constant. However, this will not significantly improve 
the width of the SC spectrum as compared to case d1, because the narrow 
direct degenerate FWM gain bands will then be in the region, where a SC 
is already generated by Raman and FWM processes. Moreover, this will lead 
to even more unusual dispersion profiles than that for case d2 (see 
Fig.~\ref{dispf}), which does not seem to be experimentally realistic.

\begin{figure}[h]
  \psfig{figure=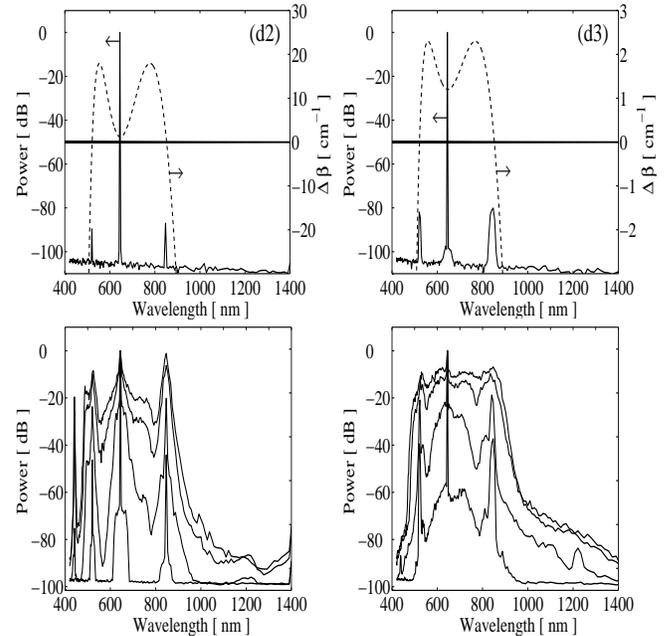,width=1\linewidth}
\caption{Dispersion profiles d2 (left) and d3 (right). Top row:
  phase mismatch $\Delta\beta$ for direct degenerate FWM 
  (dashed line) and the spectrum at $L=$10cm (solid line). 
  Bottom row: spectrum at $L$=20cm, 30cm, 1m, and 2m (down to up).}
\label{scg23}
\end{figure}

Instead we fix $\lambda_s$ and $\lambda_{as}$ while reducing the slope
and thus increasing the gain bandwidth. For dispersion profile d3 the
direct degenerate FWM gain bandwidth is thus increased to 16.5THz
(see Fig. \ref{gainf}). This leads to broader Stokes and anti-Stokes 
lines in the initial stages of the SC generation and finally to their 
merging with the spectrum around the pump as seen from Fig.~\ref{scg23}
(right). Thus a SC, which is flat within $\rm{10dB}$ and extending from 
around 500nm to 900nm is formed after a propagation distance of $L=\rm{2m}$ 
despite using low-power picosecond pulses. Moreover, the dispersion profile 
d3 is more realistic, i.e., closer to a real PCF dispersion profile, such 
as profile d1.

The SC process is thus much more efficient with dispersion profile d3
than with d1, since the power in the Stokes and anti-Stokes lines is 
not lost. However, the SC may be further improved by designing the 
dispersion such that the phase-mismatch $\Delta\beta(\lambda)$ has 
two or even three sets of Stokes and anti-Stokes lines, i.e. roots of 
the polynomium (\ref{phaseq}). 
The dispersion profiles d4 and d5 represents such cases with two and three
sets of Stokes and anti-Stokes lines, respectively, around which the 
spectrum can broaden. From the corresponding gain curves in Fig. 
\ref{gainf} we see that two gain bands actually overlap and form one 
broad gain band. The presence of extra Stokes and anti-Stokes lines 
and the broad gain band could make the SC generation more efficient, 
provided they do not deplete the pump so much that the central SC 
spectrum aroud the pump deteriorates.

\begin{figure}[h]
\psfig{figure=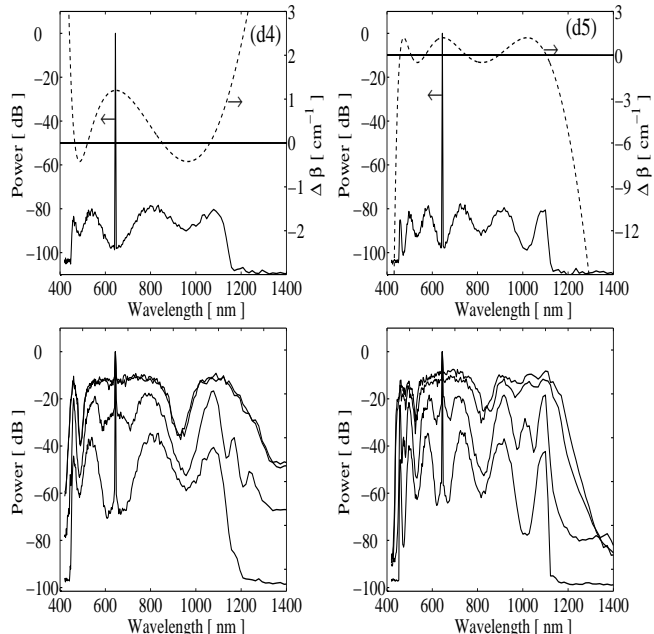,width=1\linewidth}
\caption{Dispersion profiles d4 (left) and d5 (right). Top row:
  phase mismatch $\Delta\beta$ for direct degenerate FWM 
  (dashed line) and the spectrum at $L$=10cm (solid line). 
  Bottom row: spectrum at $L$=20cm, 30cm, 1m, and 2m (down to up).}
\label{scg45}
\end{figure}

From Fig.~\ref{scg45} (left) we see that with the dispersion profile
d4 the initial stage of the SC generation is indeed improved, the 
spectrum at $L$=10cm mainly reflecting the gain profile seen in 
Fig.~\ref{gainf}. 
However, the small dip in the gain curve around 950nm has a strong 
effect on the evolution of the spectrum, leaving a clear dip at 930nm
in the final SC spectrum. Optimizing the position of the two Stokes 
lines can remove this dip and lead to an even broader SC spectrum 
than observed in Fig.~\ref{scg23} for one set of Stokes and 
anti-Stokes lines. 

Instead we show in Fig.~\ref{scg45} (right) the evolution of the 
spectrum in a PCF with the dispersion profile d5, which has three
sets of Stokes and anti-Stokes lines. The small dip in the gain 
curve around 800nm (see Fig.~\ref{gainf}) is still reflected in 
the spectrum, but it is now less pronounced and we obtain a final
ultra-broad SC spectrum ranging from 450nm to 1250nm within the 
20dB level. Of course the dispersion profile may be optimized 
further to remove the dip and make the SC spectrum more flat 
and even broader.

\begin{figure}[h]
\psfig{figure=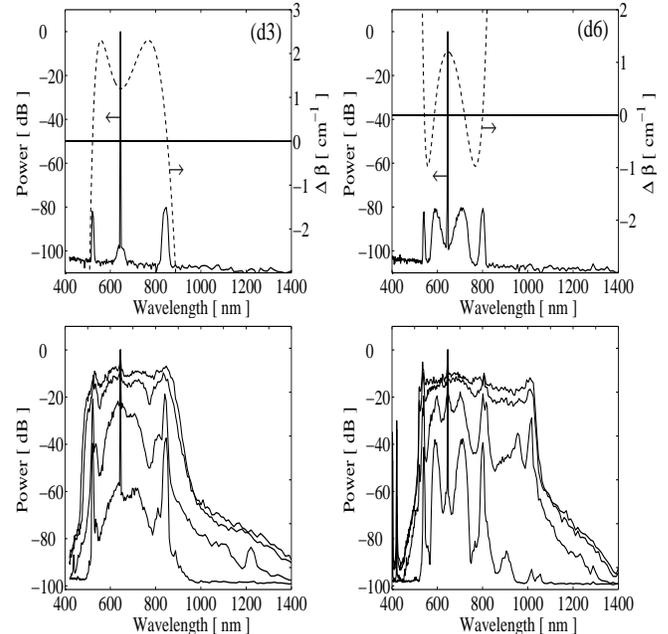,width=1\linewidth}
\caption{Dispersion profiles d3 (left) and d6 (right). Top row: 
  phase mismatch $\Delta\beta$ for direct degenerate FWM (dashed 
  line) and the spectrum at $L$=10cm (solid line). Bottom row: 
  spectrum at $L$=20cm, 30cm, 1m, and 2m (down to up).}
\label{scg36}
\end{figure}

So far we have only considered the Stokes and anti-Stokes lines generated
directly from the pump through degenerate FWM. However, so-called cascaded
FWM processes play also an important role in the evolution of the spectrum,
as discussed thoroughly in \cite{SKoen}. In particular, one can 
use these processes to obtain a broader SC. 
The dispersion profile d6 is designed to clearly illustrate this effect.
It still implies two sets of Stokes and anti-Stokes lines, but now these
are very close to the pump, within the regime of wavelengths covered by the
central SC generated by the pump. 
Nevertheless we see in Fig.~\ref{scg36} (right)
that additional lines are generated, around which the spectrum broadens,
resulting in a final SC spectrum extending from 450nm to 1$\mu$m within 
10dB. The line at 1030nm is generated by direct degenerate FWM from the 
Stokes wave at 720nm, and the line at 1060nm is generated by FWM between
the Stokes wave at 720nm and the pump. Thus these cascaded parametric
processes result in a spectrum, which is broader than what was obtained
with the direct degenerate FWM process in case d3.

Many investigations on designing the dispersion profile in PCFs have 
been made \cite{zdisp,zdisp1,zdisp2}. In \cite{zdisp1} a well-defined
procedure to design specific predetermined dispersion profiles is
established and it is shown that it is possible to obtain flattened
dispersion profiles giving normal, anomalous, and zero dispersion 
in both the telecommunication window (around 1.55$\mu$m) and the 
$Ti-Za$ laser wavelength range (around 0.8$\mu$m).
This allows us to conclude that the dispersion profiles d3-d5 shown in 
Fig. \ref{dispf} could indeed be fabricated.
However, it is outside the scope of this work to consider how the
specific dispersion profiles may be fabricated.

\section{Robustness of SCG to fiber irregularities}

The SC generation process that we have considered here is mainly 
determined by parametric FWM, which requires phase-matching. In the 
experiments with a PCF with dispersion profile d1 \cite{SKoen,SKoen1} 
Stokes and anti-Stokes lines due to direct degenerate FWM were generally 
not observed. 
This was explained to be due to irregularities along the PCF, leading 
to violation of the required phase matching condition (\ref{phaseq}). 
Thus, in order to conclude that parametric FWM can be used for broad-band 
SC generation in real PCFs, we check the robustness of the process towards 
fluctuations of the dispersion coefficients along the fiber. 

It has indeed been shown experimentally that a variation of the 
zero-dispersion wavelength $\lambda_z$ of the order of 0.1nm over the entire 
length of a dispersion shifted fiber could significantly reduce the FWM 
efficiency \cite{MKarl1}. 
This reduction in the FWM efficiency was later explained theoretically
from expressions for the average parametric gain, phase-conjugation
efficiency, and gain band-width \cite{MKarl}. 
It has also been shown that in order to control the dispersion within 
$\pm$1ps/(km$\cdot$nm), the allowable deviation of the core radius in 
W-type dispersion-flattened fibers is 0.04$\mu$m, while for other types 
of dispersion-flattened fibers the allowable core radius deviation is 
0.1$\mu$m \cite{NKuwaki}. 
As PCFs have an even more complex structure, strong fluctuation of the 
fiber dispersion could be expected too. However, to our knowledge a 
thorough study of the influence of fluctuations of the PCF parameters 
(e.g., core size and pitch) on the variation of the dispersion profile 
(i.e. the dispersion coefficients $\beta_{2-7}$) is not available.

For a newly developed highly nonlinear PCF with $\lambda_z$=1.55$\mu$m, 
it was recently shown that the variation of $\lambda_z$ is only 6nm and 
the variation of the dispersion slope at $\lambda_z$, $D_z'=dD
(\lambda_z)/d\lambda$, varies between -0.25 and -0.27 ps/(km$\cdot$nm$^2$) 
over a $150\rm{m}$ span \cite{zdisp2}. 
Expanding the propagation constant to third order around the pump 
wavelength $\lambda_p$=647nm, the dispersion has the form
\begin{equation}
   D(\lambda) = -\frac{2\pi c}{\lambda^2}\left[ \beta_2 + 2\pi c\beta_3
   \left( \frac{1}{\lambda}-\frac{1}{\lambda_p} \right) \right].
\end{equation} 
>From this expression we find the dispersion coefficients
\begin{eqnarray}
 \beta_2 = \frac{\lambda_z^4D_z'}{2\pi c} \left( \frac{1}{\lambda_p} - 
           \frac{1}{\lambda_z} \right), \quad  
 \beta_3 = \frac{\lambda_z^4D_z'}{4\pi^2c^2},
\end{eqnarray} 
which gives the extrema $\beta_2^{max}$=7.51ps$^2$/km, 
$\beta_2^{min}=6.83$ps$^2$/km, $\beta_3^{max}$=0.44ps$^3$/km, and 
$\beta_3^{min}$=0.40ps$^3$/km and thus the relative variations 
$\langle\beta_2\rangle\simeq\langle\beta_3\rangle$=9.5\%, where
\begin{equation}
 \langle\beta_k\rangle \equiv \frac{\beta_k^{max}-\beta_k^{min}}
 {(\beta_k^{max}+\beta_k^{min})/2}, \quad\quad k=1,2.
\end{equation}
Note that the relative variations of $\beta_2$ and $\beta_3$ are equal.

We model the effect of a fluctuating dispersion profile by imposing that 
$\delta\beta$, $\Delta$, and all the dispersion coefficients $\beta_{2-7}$ 
[see Eq.~(\ref{scgq})] vary randomly along the fiber,
\begin{eqnarray}
\label{ranbetaq}
 & & \delta\beta(z) = \delta\beta + \sigma_0(z),  \quad
     \beta_{k,x}(z) = \beta_k     + \sigma_{k,x}(z), \nonumber \\
 & &  \;\Delta(z)   = \;\Delta    + \sigma_1(z), \quad 
     \beta_{k,y}(z) = \beta_k     + \sigma_{k,y}(z), \nonumber
\end{eqnarray}
where $k$=2,..,7.
The random fluctuation of the coefficients, represented by the $\sigma$-terms, 
is modelled as Gaussian distributed white noise with zero mean.
To achieve the most severe case we use different seeds for all terms.
We have thus assumed that the fluctuations affect the dispersion in the 
two birefringent axis independently. 
With the results from Ref.~\cite{zdisp2} in mind we assume that the 
strength (or variance) of the fluctuations is the same in all 
coefficients,
\begin{equation}
 \frac{\langle\sigma_0^2(z)\rangle}{\delta\beta} =
 \frac{\langle\sigma_1^2(z)\rangle}{\Delta} =
 \frac{\langle\sigma_{k,x}^2(z)\rangle}{\beta_k} = 
 \frac{\langle\sigma_{k,y}^2(z)\rangle}{\beta_k} = \rho,
\label{ranrhoq}
\end{equation}
and use the strength $\rho$=10\%.

Random fluctuations of the whole dispersion profile will randomly
vary not only the zero-dispersion wavelength $\lambda_z$, but more
importantly, the phase-mismatch curve $\Delta\beta(\lambda)$, given 
by Eq.~(\ref{phaseq}). 
This in turn implies that the FWM gain spectrum $g(\Delta\beta)$, 
given by Eq.~(\ref{gainq}), will vary randomly along the fiber, even 
in the undepleted pump approximation (constant peak power $I_p$).

\begin{figure}[h]
 \centerline{\psfig{figure=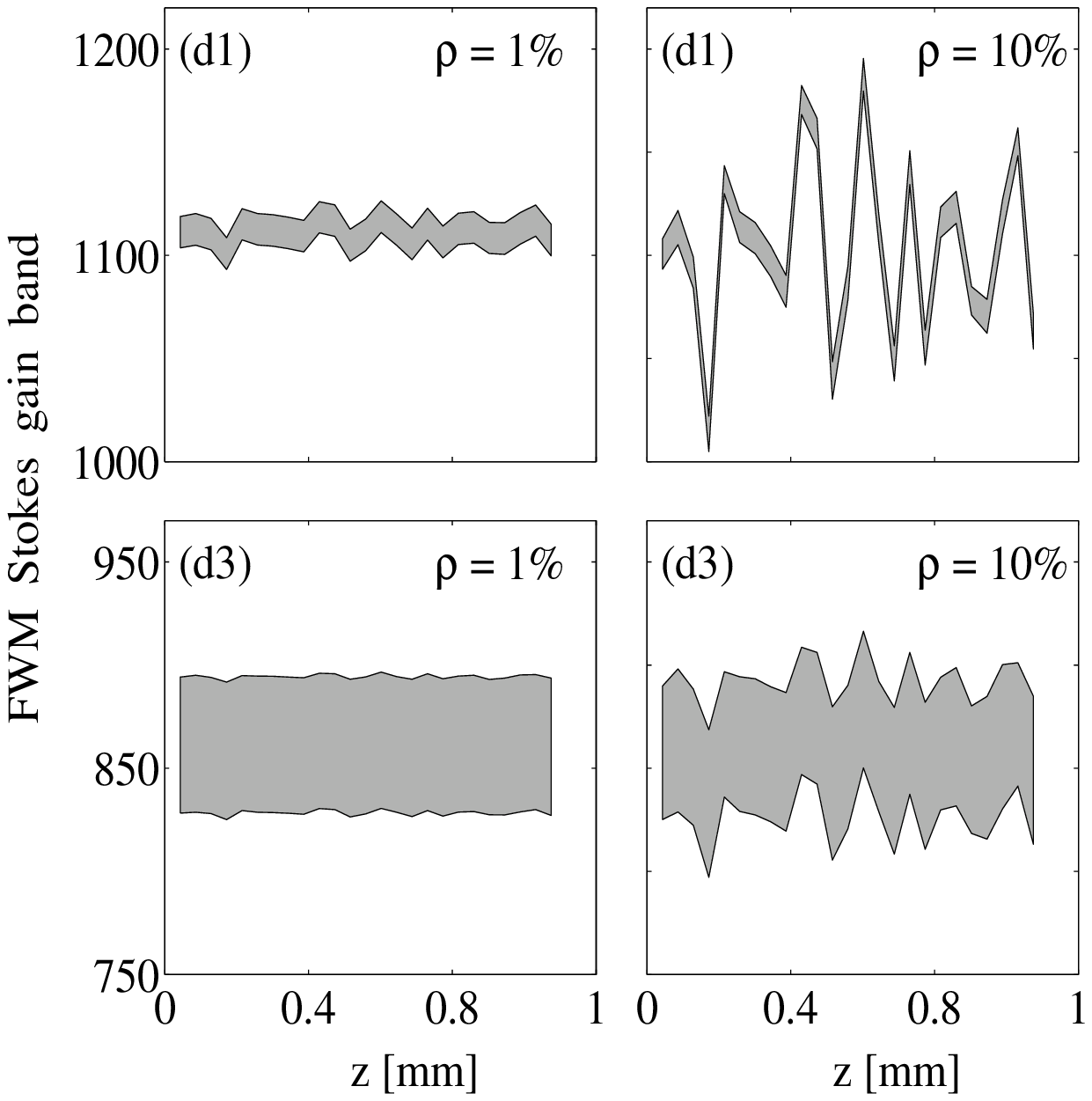,width=0.9\linewidth}}
 \caption{Random fluctuations of the FWM Stokes gain band 
 $g(\Delta\beta)$ (grey region), given by Eq.~(\ref{gainq}), 
 for constant pump $I_p$ and $\rho$=1\% (left) and $\rho$=10\% 
 (right).  The upper row is for case d1 and the bottom row for 
 case d2.}
 \label{phaseranf}
\end{figure}

In Fig.~\ref{phaseranf} we have depicted the fluctuation of the FWM
Stokes gain band in the undepleted pump approximation over the first 
$L$=1mm of a PCF with dispersion profiles d1 ($\lambda_s$=1110nm)
and d3 ($\lambda_s$=852nm).
As expected the broader gain band of fiber d3 is reflected in a
suppression of the oscillations, as compared to the fiber d1 used
in the experiments in \cite{SKoen}.

\begin{figure}[h]
  \psfig{figure=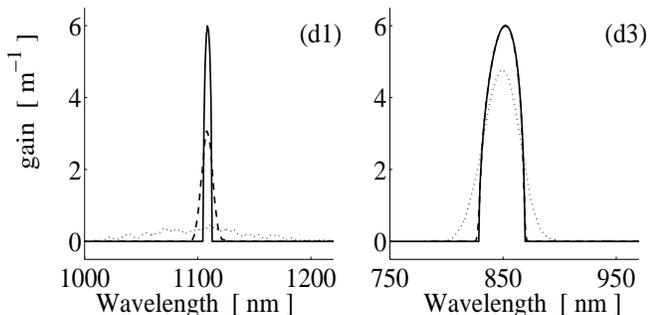,width=1\linewidth}
 \caption{Average FWM Stokes gain $g_{av}$ over $L$=2cm, as 
 given by Eq.~(\ref{avgainq}) in the undepleted pump approximation.
 Shown is case d1 (left) and d3 (right) for $\rho$=0 (solid),
 $\rho$=1\% (dashed), and $\rho$=10\% (dotted).}
 \label{avgainf}
\end{figure}

The variation of the direct degenerate FWM Stokes gain band gives an
impression of the influence of the dispersion fluctuations on the 
effectiveness of the FWM process. 
The important factor is the average FWM gain over the fiber length
$L$, defined as
\begin{equation}
   g_{av} \equiv \frac{1}{L} \int_0^L g[\Delta\beta(z)]~dz,
   \label{avgainq}
\end{equation}
where we have indicated the $z$-dependence of the phase-mismatch as 
a result of the fluctuations.
In Fig.~\ref{avgainf} we show the average FWM Stokes gain calculated
over the first $L$=2cm using the undepleted pump approximation.
The reduction of the average gain for increasing strength of the 
fluctuations can be clearly observed.

\begin{figure}[h]
\psfig{figure=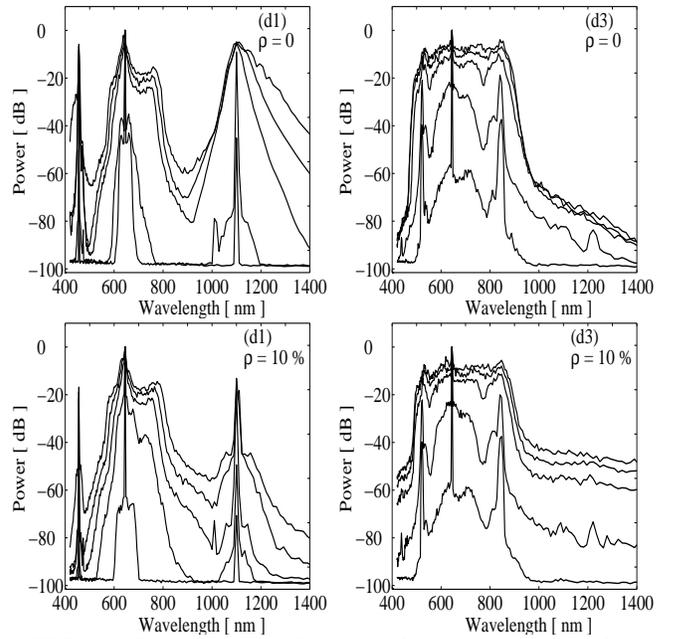,width=1\linewidth}
\caption{Dispersion profiles d1 (left) and d3 (right). Spectrum under
  influence of fluctuations with strength $\rho$=0 (top row) and
  $\rho$=10\% (bottom row) at $L$=20cm, 30cm, 1m, 2m, and 3.7m (down to up).}
\label{scgranf}
\end{figure}

Theoretically we thus predict that in fiber d1 realistic 
fluctuations would significantly suppress the Stokes and anti-Stokes 
lines generated from the pump by direct degenerate FWM, as also 
stated by Coen {\it et al.} \cite{SKoen}.
The corresponding simulations, presented in Fig.~\ref{scgranf} (left),
confirms this prediction.
Using direct degenerate FWM to generate an ultra-broad SC in the
particular fiber d1 is therefore not realistic.

In contrast, with our proposed fiber d3 with a broad gain band, even
fluctuations with $\rho$=10\% should not significantly reduce the
FWM effectiveness, which is also confirmed by our simulations
shown in Fig.~\ref{scgranf} (right).
In our proposed fibers d4-d5 the FWM gain band is even broader,
indicating that fluctuations will have even less impact.
Thus our numerical results show that using direct degenerate FWM 
to generate an ultra-broad SC in really a realistic option.

\section{Conclusion}

We have numerically investigated SC generation in birefringent PCFs using
30-picosecond pulses with 400-kilowatt peak power. 
Our results show that by properly 
designing the dispersion profile and using the generation, broadening, 
and final merging of widely separated pump and FWM Stokes
and anti-Stokes lines the SC generation efficiency can be significantly
improved, resulting in a broader SC spectrum and a reduced loss of power
to frequencies outside the SC.
Thus, by optimising the dispersion profile, we have generated an 
ultra-broad SC ranging from 450nm to 1250nm within the 20dB level.

We have shown that the key issue is to make sure that the Stokes and
anti-Stokes lines are generated close enough to the pump to be able 
to broaden and merge with the central (pump) part of the SC before
nonlinear processes are suppressed due to fiber loss and temporal 
walk-off. We have also shown that this in turn requires the FWM
gain band to be sufficiently broad, which is an essential property 
of our designed dispersion profiles.

We have finally investigated the robustness of the SC generation process 
in our proposed fibers towards fluctuations in the parameters along the 
fiber. 
Such fluctuation could be detrimental to the phase-sensitive FWM
process, which depends on the degree of phase-matching.
Simulations including random fluctuations of the dispersion profile 
along the fiber show that the broad FWM gain band of our proposed 
fibers improve the robustness and that the process of efficient SC 
generation survives random fluctuations of a realistic strength.

This work was supported by the Danish Technical Research Council (Grant
no. 26-00-0355) and the Graduate School in Nonlinear Science (The
Danish Research Agency).

\end{multicols}  

\end{document}